\documentclass[runningheads]{llncs}
\usepackage[T1]{fontenc}
\usepackage[utf8]{inputenc}
\usepackage{amsmath,amssymb,amsfonts,bm}
\usepackage{graphicx,tabularx}
\usepackage{microtype}
\usepackage{lmodern}
\usepackage[english]{babel}
\usepackage[bottom]{footmisc}
\usepackage{booktabs}
\usepackage{fontawesome5}
\usepackage[hidelinks]{hyperref}
\usepackage{tikz}
\usetikzlibrary{positioning, fit, backgrounds, arrows.meta, shapes.geometric, calc, shadows}

\begin{document}

\title{Accelerating Dynamic Graph Clustering \\ on GPU Architectures with cuGraph}
\titlerunning{Accelerating Dynamic Graph Clustering on GPUs}

\author{
    Nelson Aloysio Reis de Almeida Passos\inst{1,2}\orcidID{0000-0003-1869-2976} \and
    Emanuele Carlini\inst{2}\orcidID{0000-0003-3643-5404} \and
    Salvatore Trani\inst{2}\orcidID{0000-0001-6541-9409}
}
\authorrunning{N. A. R. A. Passos et al.}

\institute{
    University of Pisa, Dept. of Computer Science, 56127 Pisa PI, Italy \and
    National Research Council, 56124 Pisa PI, Italy \\[1em]
    \email{
        nelson.reis@phd.unipi.it \\
        \{emanuele.carlini,salvatore.trani\}@isti.cnr.it}}

\maketitle

\vspace{-1em}
\begin{abstract}
This work addresses community detection in temporal networks through GPU-accelerated extensions of spectral clustering and modularity-based algorithms originally designed for static graphs.
Built on the NVIDIA RAPIDS ecosystem, the framework enables the characterization and tracking of communities in snapshot-based dynamic graphs, either by Leiden greedy optimization with multi-GPU support via Dask-based workload distribution, or eigendecomposition of a symmetric Bethe-Hessian operator.
Our multislice modularity backend achieves up to roughly three orders of magnitude speedup over the CPU reference under an equal-work budget, depending on graph density and snapshot count, while preserving compatibility with existing graph analytics pipelines.
We demonstrate its applicability on real-world and synthetic datasets, facilitating exploratory analysis of structural network properties over time.
Such capabilities are relevant across several application domains, such as epidemic spreading, financial systems, cybersecurity, and trajectory and mobility analysis.
We release our implementation as free and open-source software, including Python bindings through the NetworkX-Temporal\footnote{
    \url{https://www.networkx-temporal.org}. Distributed under the 3-Clause BSD License.
} library for ease of use and zero-code acceleration with existing codebases.

\end{abstract}

\section{Introduction} \label{sec:intro}

Community detection is a fundamental task in network science, which seeks to identify groups of nodes with functional similarity or shared characteristics.
The identification of functional modules in cell structures, the detection of botnets and malicious actors in computer networks, and the discovery of users in social media platforms with shared political or cultural interests are just a few examples of its many applications in real-world settings.
Among the many techniques employed for the task, there are inference-based statistical models relying on sampling and optimization strategies, such as Markov-Chain Monte Carlo and simulated annealing; matrix factorization based on eigendecomposition or random walk sampling; heuristic algorithms optimizing some sort of `quality' function; and more recently, neural models that map nodes to a real-valued vector space, where distances encode structural or attribute similarity.
Each method carries its own advantages and trade-offs, and their suitability is expected to be highly dependent on the specific context and requirements of the problem at hand \cite{fortunato2016,peixoto2023}.

By allowing nodes and edges to change over time, dynamic (temporal) graphs \cite{temporalnetworktheory2023} offer a powerful mathematical framework for the analysis of non-Euclidean relational data, where individual elements (nodes) and their connections (edges) may fluctuate in response to external stimuli and internal dynamics.
This enables the modeling of evolving relationships and interactions, crucial for studying real-world phenomena such as information diffusion, epidemic spreading, financial fraud, and traffic analysis.
Such flexibility, however, also brings additional burdens, for both the definition of communities themselves (i.e., their identity) and the algorithmic implementations for their detection \cite{fortunato2016}, which must now account for these additional dynamics to ensure their consistent tracking across time.

In this work, we address the challenge of efficiently managing the increased complexity arising from node clustering in dynamic graphs by implementing GPU-based (graphics processing units) versions of widely adopted techniques for the task, with analytical tractability and scalability as primary goals.
The analysis of large-scale networks exemplifies an area where GPUs can deliver a superior performance-to-cost ratio over CPU-bound operations due to higher memory bandwidth and massive parallelism, motivating the present research.
We focus on three algorithms along two pathways for temporal community detection, relying on graph spectra and heuristic optimization: eigendecomposition and clustering of either the Bethe-Hessian \cite{bethehessian_saade2014} or the modularity matrix \cite{spectral_modularity_newman2006}; and Leiden optimization of multislice modularity \cite{modularity_mucha2010,leiden_traag2019}, which maximizes the homonymous quality function measuring the density of connections within communities compared to a null model, i.e., a random (`configuration') graph with the same degree distribution as the observed network \cite{chunglu2002}.
Our specific contributions are that we:
\begin{itemize}
    \setlength\itemsep{-0.8em}
    \item Implement three algorithms along two pathways for temporal community detection, using the open-source GPU libraries cuGraph for modularity optimization and CuPy with cuML for spectral clustering \cite{cugraph_docs,cuml_docs,cupy_learningsys2017};\\
    \item Demonstrate that they enable the characterization and tracking of evolving communities, with the Leiden backend allowing multi-GPU execution via Dask \cite{dask_docs} and achieving substantial speedups over its CPU reference; and\\
    \item Integrate our implementations into the open-source NetworkX-Temporal library \cite{networkxtemporal2025}, enhancing its functionality and promoting user accessibility through zero-code-change GPU acceleration of existing workflows.
\end{itemize}

The remainder of this paper is as follows.
Section 2 discusses related work on static and temporal community detection.
Section 3 details our methodology, including the algorithms we extend and their temporal formulations.
Section 4 presents experimental results on several benchmark datasets.
This work concludes by summarizing our contributions, limitations, and future research directions.

\section{Background \& Related Work} \label{sec:rw}

The definition, identification, and tracking of communities in dynamic graphs remain open problems, compounded by the absence of a universally accepted notion of what constitutes a community and the inherent complexity of temporal networks \cite{fortunato2016,temporalnetworktheory2023}.
The most common definition of community is that of a group of nodes with dense internal connections compared to connections to other groups, therefore understood as an \textit{assortative} structure.
Based on an assumption of homophily --- `birds of a feather flock together' \cite{mcpherson2001} --- it posits that nodes with similar features or connectivity patterns are more likely to belong to the same group, and is the prevailing definition in tasks such as social network analysis, where individuals with similar interests or characteristics tend to form communities.
Meanwhile, other systems may exhibit different characteristics, such as hierarchical (e.g., nested communities), disassortative (bipartite), core-periphery, or overlapping (mixed-membership) structures, which call for alternative strategies and definitions outside the scope of this work \cite{fortunato2016}.

This work focuses on modularity optimization and spectral clustering, two techniques widely adopted for community detection in static graphs, extensible to the temporal domain and suitable for GPU-based implementations.
Both are commonly employed due to their relative ease of implementation, strong empirical performance, and compatibility with sparse matrices, and traditionally rely on CPU-bound implementations designed for static graphs.
Other solutions include divisive strategies for iterative edge removal, such as edge betweenness centrality \cite{community_girvan2002}; statistical physics approaches formulating community detection as an energy minimization problem, including Ising- and Potts-based models \cite{community_reichardt2004,community_son2006}; inferential models based on, e.g., stochastic block modeling (SBM) \cite{sbm_peixoto2019}; information-theoretic techniques, such as Infomap \cite{infomap_rosvall2008}; propagation-based methods that iteratively diffuse labels across local neighborhoods \cite{raghavan2007}; non-negative matrix factorization \cite{survey_he2022} and matrix decomposition by random walk sampling \cite{walktrap_pons2006}; graph embedding methods to obtain low-dimensional node representations, also based on walk sampling \cite{survey_zhang2021}; hybrid and ensemble approaches combining multiple strategies \cite{yan2019}; and more recent deep (multilayer) neural network-based models with strong empirical performance that approximate (`learn') a function mapping nodes $\mathcal{V}$ of a graph $\mathcal{G}$ into a real, $d$-dimensional (latent) space --- where distances among nodes encode their similarity based on both the system structure (topology) defined by edges $\mathcal{E}$, and high-dimensional node attributes $\mathcal{X}$, i.e., $f(\mathcal{G}) \to \mathbb{R}^d$, where $\mathcal{G} = (\mathcal{V}, \mathcal{E}, \mathcal{X})$ --- either end-to-end differentiable or, much more frequently, subsequently clustered using conventional methods such as $k$-means \cite{survey_su2022}.

This work focuses instead on non-attributed, large-scale dynamic graph data.
Although several methods have been extended to the temporal domain, e.g., by segmenting networks into slices (snapshots) and employing strategies like dynamic SBM, inter-slice couplings, and dynamic embedding procedures \cite{sbm_cugmas2023,modularity_mucha2010,survey_xue2022}, they often rely on expensive CPU-bound operations, such as repeated inference, supra-graph construction, or orthogonal matrix alignment, which hinder their applicability at scale or under high temporal resolution, including mobility and trajectory analysis, social network monitoring, and cybersecurity applications.

\section{Methodology \& Implementation} \label{sec:methodology}

This section details the algorithms we extend and their temporal formulations, along with implementation details and integration into the NetworkX-Temporal library \cite{networkxtemporal2025}.
We define spectral clustering on supra-graphs and multislice (temporal) modularity optimization, the latter of which is evaluated at scale in Section~\ref{sec:results}.

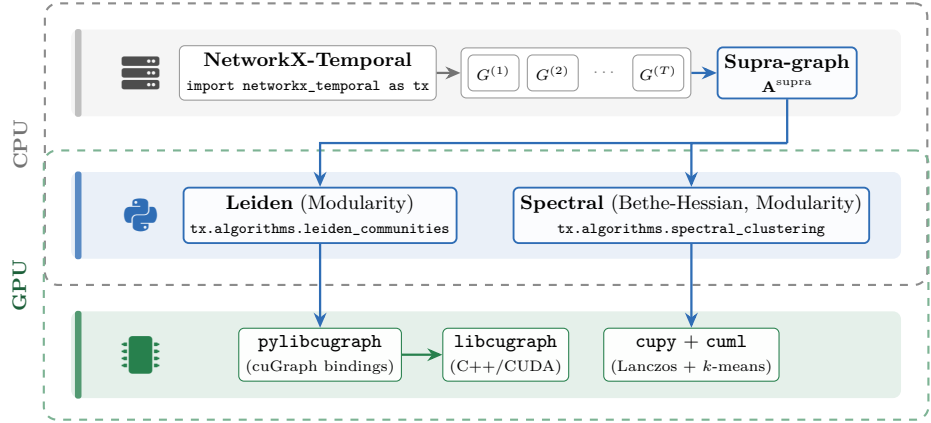
\begin{figure}[b!]
\centering
\begin{tikzpicture}[
    font=\small,
    scale=0.82,
    transform shape,
    band/.style={
        inner sep=0pt,
        rounded corners=3pt,
    },
    component/.style={
        align=center,
        draw=black!40,
        fill=white,
        font=\footnotesize,
        minimum height=0.8cm,
        minimum width=2.3cm,
        rounded corners=2pt,
    },
    data/.style={
        align=center,
        draw=black!40,
        fill=white,
        font=\footnotesize,
        minimum height=0.8cm,
        minimum width=1.7cm,
        rounded corners=2pt,
    },
    graph/.style={
        draw=black!40,
        fill=white,
        font=\scriptsize,
        inner sep=1.8pt,
        minimum height=0.58cm,
        minimum width=0.82cm,
        rounded corners=2pt,
    },
    snapshots/.style={
        draw=black!40,
        rounded corners=2pt,
        inner sep=2.5pt,
    },
    contribcomp/.style={
        align=center,
        draw=blue,
        fill=blue!2,
        font=\footnotesize,
        line width=0.7pt,
        minimum height=0.8cm,
        minimum width=2.3cm,
        rounded corners=2pt,
    },
    flow/.style={-{Stealth[length=2.0mm, width=1.7mm]}, draw=black!55, line width=0.85pt},
    contribflow/.style={-{Stealth[length=2.0mm, width=1.7mm]}, draw=blue, line width=0.85pt},
    gpuflow/.style={-{Stealth[length=2.0mm, width=1.7mm]}, draw=green, line width=0.85pt},
    sidelabel/.style={font=\footnotesize\bfseries, rotate=90},
    node distance=3.5mm
]

\definecolor{blue}{HTML}{2F6DB5}
\definecolor{green}{HTML}{27804D}
\def\bandHalfHeight{7mm}

\coordinate (iconcol) at (0,0);
\node[font=\large\color{black!70}] (cpuicon) at (iconcol) {\LARGE \faServer};
\node[component, right=2mm of cpuicon] (api)
    {\textbf{NetworkX-Temporal}\\\texttt{\scriptsize import networkx\_temporal as tx}};

\node[graph, right=5mm of api] (s1) {$G^{(1)}$};
\node[graph, right=1.2mm of s1] (s2) {$G^{(2)}$};
\node[right=1.2mm of s2, font=\scriptsize\color{black!50}] (s3) {$\cdots$};
\node[graph, right=1.2mm of s3] (sT) {$G^{(T)}$};
\node[snapshots, fit=(s1)(sT)] (snapgroup) {};
\node[data, draw=blue, line width=0.7pt, fill=blue!2, right=4.1mm of snapgroup] (supra)
    {\textbf{Supra-graph}\\\scriptsize{$\mathbf{A}^\mathrm{supra}$}};

\coordinate (bandL) at ([xshift=-7mm]cpuicon.west);
\coordinate (bandR) at ([xshift=7mm]supra.east);
\coordinate (topNW) at ([yshift=\bandHalfHeight]bandL);
\coordinate (topSE) at ([yshift=-\bandHalfHeight]bandR);

\begin{scope}[on background layer]
    \node[band, fill=black!4, fit=(topNW)(topSE)] (topband) {};
    \fill[black!25, rounded corners=1pt]
        ([xshift=2pt]topband.north west) rectangle ([xshift=5pt]topband.south west);
    \node[draw=none, fill=white, rounded corners=2pt, inner sep=2.5pt, fit=(s1)(sT)] (snapgroupbg) {};
\end{scope}

\coordinate (midL) at ([yshift=-2.30cm]bandL);
\coordinate (midR) at (bandR |- midL);
\coordinate (midNW) at ([yshift=\bandHalfHeight]midL);
\coordinate (midSE) at ([yshift=-\bandHalfHeight]midR);

\node[contribcomp, anchor=west] (leiden) at ([xshift=18mm]midL)
    {\textbf{Leiden} \small{(Modularity)} \\\scriptsize{\texttt{tx.algorithms.leiden\_communities}}};
\node[contribcomp, anchor=east] (spectral) at ([xshift=-5mm]midR)
    {\textbf{Spectral} \small{(Bethe-Hessian, Modularity)} \\\scriptsize{\texttt{tx.algorithms.spectral\_clustering}}};
\node[font=\large\color{blue}] (pyicon) at (iconcol |- midL) {\LARGE \faPython};

\begin{scope}[on background layer]
    \node[band, fill=blue!10, fit=(midNW)(midSE)] (contribband) {};
    \fill[blue!80, rounded corners=1pt]
        ([xshift=2pt]contribband.north west) rectangle ([xshift=5pt]contribband.south west);
\end{scope}

\coordinate (botL) at ([yshift=-4.55cm]bandL);
\coordinate (botR) at (bandR |- botL);
\coordinate (botNW) at ([yshift=\bandHalfHeight]botL);
\coordinate (botSE) at ([yshift=-\bandHalfHeight]botR);

\node[component, draw=green, fill=green!1, minimum width=2cm] (plc) at (leiden.south |- botL) {\texttt{pylibcugraph} \\ \scriptsize {(cuGraph bindings)}};
\node[component, draw=green, fill=green!1, minimum width=2cm] (cupy) at (spectral.south |- botL) {\texttt{cupy} + \texttt{cuml} \\ \scriptsize{(Lanczos + $k$-means)}};
\node[component, draw=green, fill=green!1, minimum width=2cm] (libcugraph) at ($(plc)!0.5!(cupy)$) {\texttt{libcugraph}\\ \scriptsize{(C++/CUDA)}};
\node[font=\large\color{green}] (gpuicon) at (iconcol |- botL) {\LARGE \faMicrochip};

\begin{scope}[on background layer]
    \node[band, fill=green!12, fit=(botNW)(botSE)] (rapidsband) {};
    \fill[green!80, rounded corners=1pt]
        ([xshift=2pt]rapidsband.north west) rectangle ([xshift=5pt]rapidsband.south west);
\end{scope}

\begin{scope}[on background layer]
    \node[draw=black!45, dashed, line width=0.8pt, rounded corners=5pt,
          inner sep=3.4mm,
          fit=(topband)(contribband)] (pkgbox) {};
    \node[sidelabel, color=black!55, anchor=south] at ([xshift=-1.5mm]pkgbox.west) {CPU};
\end{scope}

\begin{scope}[on background layer]
    \coordinate (rtop) at ([yshift=3.4mm]contribband.north);
    \coordinate (rbot) at ([yshift=-3.4mm]rapidsband.south);
    \node[draw=green!60, dashed, line width=0.8pt, rounded corners=5pt,
          inner sep=0pt,
          fit={(pkgbox.west |- rtop) (pkgbox.east |- rbot)}] (rapidsbox) {};
    \node[sidelabel, color=green!75!black, anchor=south] at ([xshift=-1.5mm]rapidsbox.west) {GPU};
\end{scope}

\draw[flow] (api.east) -- (snapgroup.west);
\draw[contribflow] (snapgroup.east) -- (supra.west);

\coordinate (fork) at ($(supra.south)!0.5!(supra.south|-leiden.north)$);
\draw[contribflow,-] (supra.south) -- (fork);
\draw[contribflow] (fork) -| (leiden.north);
\draw[contribflow] (fork) -| (spectral.north);

\draw[contribflow] (leiden.south) -- (plc.north);
\draw[contribflow] (spectral.south) -- (cupy.north);

\draw[gpuflow] (plc.east) -- (libcugraph.west);

\end{tikzpicture}
\caption{
    \textbf{NetworkX-Temporal integration.}
    Temporal graph snapshots $G^{(t)}$ are assembled into a supra-graph and dispatched through one of two GPU-accelerated paths: multislice modularity via Leiden optimization, or spectral clustering of a (temporal) Bethe-Hessian or modularity matrix.
    All methods execute on the RAPIDS ecosystem and return per-snapshot community assignments, allowing zero-code-change execution and device selection by setting an environment variable.
    In blue: our contributions.
    }
\label{fig:architecture}
\end{figure}

\subsection{Spectral Clustering} \label{sec:spectral}

Spectral clustering relies on the eigendecomposition of graph operators to obtain low-dimensional node embeddings that capture the global structure of the network.
Specifically, spectral methods cluster vertices by eigendecomposition of the normalized Laplacian $\tilde{\mathbf{L}} = \mathbf{D}^{-1/2}\mathbf{L}\mathbf{D}^{-1/2}$, where $\mathbf{L}=\mathbf{D}-\mathbf{A}$ is the graph Laplacian, $\mathbf{D}$ is the degree matrix, and $\mathbf{A}$ is the adjacency matrix.
The $k$ leading eigenvectors of $\tilde{\mathbf{L}}$ form a matrix $\mathbf{U}\in\mathbb{R}^{n \times k}$, where each row corresponds to a node embedding; these rows are then clustered using $k$-means or another partitioning algorithm \cite{spectral_ng2002}.
The second smallest eigenvector, known as the Fiedler vector, can be used to bipartition a graph by thresholding its values, and its relation to graph connectivity is formalized by Cheeger's inequality, which bounds the conductance of the best partition by the second smallest eigenvalue of $\tilde{\mathbf{L}}$ \cite{chung1996}.

The choice of objective function determines which Laplacian is used and how the resulting partition is evaluated.
For instance, minimizing the normalized cut \cite{spectral_shi2000} encourages partitions with dense intra-community and sparse inter-community connections, i.e., normalized by edge volume. It is expressed as
\begin{equation}
    \label{eq:ncut}
    \text{NCut}(C_1,\dots,C_k) = \sum_{i=1}^{k}\frac{\text{cut}(C_i,\bar{C}_i)}{\text{vol}(C_i)},
    \quad \text{cut}(C_i,\bar{C}_i) = \sum_{u\in C_i,\,v\notin C_i}A_{uv},
\end{equation}
where $\text{vol}(C_i)=\sum_{u\in C_i}d_u$ is the sum of node degrees in $C_i$, and $\text{cut}(C_i,\bar{C}_i)$ is the total volume of edges connecting nodes in $C_i$ to those outside of it.
A similar formulation\footnote{
    \label{fn:normalization}
    Symmetric normalization introduces a $\mathbf{D}^{1/2}$ scaling that requires an additional (heuristic) row-normalization step to correct for high node degree heterogeneity \cite{spectral_vonluxburg2007}.
} employs the random-walk Laplacian $\mathbf{\hat{L}} = \mathbf{D}^{-1}\mathbf{L} = \mathbf{I} - \mathbf{D}^{-1}\mathbf{A}$, which shares the eigenvalues of $\tilde{\mathbf{L}}$, but whose eigenvectors correspond directly to the relaxed cluster indicator vectors, solvable as a generalized eigenproblem.

Minimizing balanced ratio cut \cite{spectral_hagen1992} instead considers partition size $|C_i|$ rather than edge volume, and is the default implementation in cuGraph \cite{cugraph_docs}, defined as
\begin{equation}
    \label{eq:rcut}
    \text{RCut}(C_1,\dots,C_k) = \sum_{i=1}^{k}\frac{\text{cut}(C_i,\bar{C}_i)}{|C_i|}.
\end{equation}
Note that, while both objectives are NP-hard in their discrete form, their continuous relaxations allow efficient approximations corresponding to eigenvector problems on the normalized ($\tilde{\mathbf{L}}$) and unnormalized ($\mathbf{L}$) graph Laplacians, which can be solved using iterative methods such as the Lanczos algorithm or power iteration, efficient for large sparse graphs.
Spectral methods are computationally efficient on sparse graphs, but require the number of communities $k$ to be specified in advance and are sensitive to disconnected graph components, i.e, groups of nodes with no paths in between, and may therefore yield suboptimal partitions on dynamic graphs where nodes are not present across all time steps.

Constructing a supra-Laplacian $\mathbf{L}^\mathrm{supra}$ from (temporal) adjacencies $\mathbf{A}^\mathrm{supra}$ encoding intra-slice connections and inter-slice couplings as off-diagonal blocks allows for spectral clustering techniques to be extended to the temporal domain,
\setlength{\arraycolsep}{6pt}
\begin{equation}
    \label{eq:supra-adjacency}
    \mathbf{A}^\mathrm{supra} =
    \left(
    \begin{array}{cccc}
        \mathbf{A}^{(1)} & \omega \mathbf{I} & \cdots & \mathbf{0} \\
        \omega \mathbf{I} & \mathbf{A}^{(2)} & \cdots & \mathbf{0} \\
        \vdots & \vdots & \ddots & \vdots \\
        \mathbf{0} & \mathbf{0} & \cdots & \mathbf{A}^{(T)}
    \end{array}
    \right),
\end{equation}
where $\mathbf{A}^{(t)}$ is the adjacency matrix of snapshot $t$, $\omega$ is the inter-slice coupling strength that encourages temporal persistence of community assignments across snapshots, and $\mathbf{I}$ is the identity matrix connecting corresponding temporal nodes.

Although this enables the use of iterative solvers (Lanczos, power iteration), leveraging GPU acceleration for eigendecomposition and clustering, the resulting matrix has a prohibitive order of $nT \times nT$, where $n$ is the number of nodes and $T$ is the number of snapshots.
Moreover, it is well-known that Laplacian spectral methods fail in sparse graphs near the detectability threshold of communities, i.e., the regime where many real temporal networks are expected to lie.
To escape these issues, we first consider instead a principled alternative based on linearizing belief propagation \cite{detectability_ghasemian2016}, yielding the non-backtracking (Hashimoto-type) operator
\setlength{\arraycolsep}{6pt}
\begin{equation}
    \label{eq:nonbacktracking}
    \mathbf{B} =
    \left(
    \begin{array}{cccc}
        \lambda\mathbf{A}^\mathrm{s} & -\lambda\mathbf{I} & \lambda\mathbf{A}^\mathrm{s} & \;\mathbf{0} \\
        \lambda(\mathbf{D}^\mathrm{s} - \mathbf{I}) & \;\mathbf{0} & \lambda\mathbf{D}^\mathrm{s} & \;\mathbf{0} \\
        \eta\mathbf{A}^\mathrm{t} & \;\mathbf{0} & \eta\mathbf{A}^\mathrm{t} & -\eta\mathbf{I} \\
        \eta\mathbf{D}^\mathrm{t} & \;\mathbf{0} & \eta(\mathbf{D}^\mathrm{t} - \mathbf{I}) & \;\mathbf{0}
    \end{array}
    \right),
\end{equation}
where each of $\mathbf{A}^\mathrm{s}$, $\mathbf{A}^\mathrm{t}$, $\mathbf{D}^\mathrm{s}$, $\mathbf{D}^\mathrm{t}$ is an $nT \times nT$ matrix: $\mathbf{A}^\mathrm{s} = \bigoplus_t \mathbf{A}^{(t)}$ is the block-diagonal intra-slice adjacency matrix, $\mathbf{A}^\mathrm{t}$ is the inter-slice adjacency matrix connecting each node to its time-adjacent copies with block-bidiagonal identities, $\mathbf{D}^\mathrm{s}$ and $\mathbf{D}^\mathrm{t}$ are the corresponding diagonal degree matrices, and $\mathbf{I}$ is the $nT$-dimensional identity.
The scalar $\lambda = (c_\mathrm{in} - c_\mathrm{out})/(k\,c)$ is the second eigenvalue of the (normalized) $k \times k$ block affinity matrix --- with $c_\mathrm{in}$ and $c_\mathrm{out}$ the average within- and between-community connection rates, $k$ the number of communities, and $c$ the average degree --- and governs how storngly community signal propagates along spatial edges; analogously, $\eta$ governs its propagation along temporal edges.
Together they yield a more flexible and expressive formulation than that based solely on $\omega$ (Eq.~\ref{eq:supra-adjacency}), with effective SNR $c\lambda^2$ dictating detectability\footnote{
    Below the information-theoretic threshold (Almeida--Thouless line), recovery is impossible for any algorithm, while $c\lambda^2=1$ (Kesten--Stigum threshold) defines the limit of efficient recovery for polynomial-time algorithms \cite{detectability_ghasemian2016}. For $k>4$ communities, they separate, defining a regime where detection is possible but not known to be efficient.
} and $\eta \in [0, 1]$ the probability that a node retains its community label on consecutive time steps.

The construction is intentionally asymmetric (direction-aware), yielding a method that preserves the arrow of time, is asymptotically optimal in the sparse regime \cite{detectability_ghasemian2016}, and robust to degree heterogeneity, avoiding the localization on high-degree nodes of Laplacian operators (Footnote~\ref{fn:normalization}).
Although the resulting matrix is $4nT \times 4nT$, it is also critically sparse and therefore more amenable to iterative solvers on GPUs:
each block of $\mathbf{B}$ corresponds to either an adjacency or diagonal matrix, and its $k-1$ eigenvectors with largest eigenvalues may be extracted to perform $k$-means clustering on the resulting $n$ vectors in $\mathbb{R}^{k-1}$.

However, sparse GPU eigensolvers are currently restricted\footnote{
    Although ARPACK's non-symmetric Arnoldi routines are available in SciPy, their use on GPUs requires matvec callbacks (CPU-side iterations of per-step host-device transfers), effectively negating the potential gains of acceleration for large-scale problems.
} to symmetric or Hermitian matrices, so the asymmetric non-backtracking operator $\mathbf{B}$ cannot be solved directly by available routines.
We therefore reformulate it as a symmetric problem by considering the Bethe-Hessian $\mathbf{H}(r) = (r^2 - 1)\mathbf{I} - r\mathbf{A} + \mathbf{D}$ \cite{bethehessian_saade2014}, where $\mathbf{A}$ and $\mathbf{D}$ are the adjacency and degree matrices of the supra-graph, and $r$ is a regularization parameter.
This maps the informative (out-of-bulk) eigenvalues of $\mathbf{B}$ to the negative eigenvalues of $\mathbf{H}(r)$, effectively preserving asymptotic optimality in the static case with appropriate $r$ selection.
If $r=1$, $\mathbf{H}(r)$ reduces to the standard Laplacian $\mathbf{L}$; in the static SBM regime, $r$ is optimally set near $\sqrt{c}$, i.e., a degree-based approximation (refined in our implementation for degree heterogeneity), while temporal extensions require accounting for both spatial and temporal connectivity, characterized only in specific settings \cite{dallamico2020}.
This symmetric eigenproblem is solved fully on-GPU using open-source libraries \cite{cuml_docs,cupy_learningsys2017} --- yielding an efficient and consistent algorithm \cite{zhao2012} for clustering on dynamic graphs.

\subsection{Modularity Optimization} \label{sec:modularity}

Modularity $Q$ aims to quantify the extent to which a network exhibits community structures by comparing the density of intra-community edges against a random graph that preserves the degree distribution of the observed network \cite{chunglu2002}.
Let $m$ be the number of edges, $d_i^{\text{out}}$ and $d_j^{\text{in}}$ the out- and in-degrees of nodes $i$ and $j$, $c_i$ the community of node $i$, and $\gamma = 1$ (default) the resolution parameter. Then,
\begin{equation}
    \label{eq:modularity}
    Q = \frac{1}{2m}\sum_{i,j}
    \Bigl(A_{ij} - \gamma \frac{d^{\text{out}}_{i}\,d^{\text{in}}_{j}}{2m}\Bigr)\,
    \delta(c_i,c_j),
\end{equation}
where $\delta(c_i,c_j) = 1$ if nodes $i$ and $j$ belong to the same community and $0$ otherwise.
Higher values of $Q$ indicate denser intra-community and sparser inter-community connections relative to the null model; while $\gamma$ allows tuning the resolution\footnote{
    Modularity maximization is subject to a resolution limit \cite{fortunato2007}: because the null model term couples all node pairs globally, communities smaller than a scale determined by $\sqrt{2m}$ may be merged into larger ones, even in the partition that globally maximizes $Q$.
} of detected communities, with higher values favoring smaller community sizes.

Maximizing $Q$ over all possible partitions is NP-hard \cite{fortunato2016}, so heuristic approaches are employed in practice.
The Leiden algorithm \cite{leiden_traag2019} adds a refinement phase to the widely used Louvain method \cite{louvain_blondel2008}, which iteratively merges nodes into communities to maximize local modularity gains and recursively aggregates communities into supernodes --- resulting in faster runtimes and guaranteeing well-connected partitions, a built-in modeling assumption.
The heuristic extends to directed and weighted graphs, and supports alternative quality functions such as surprise \cite{surprise_aldecoa2013} and the Constant Potts Model (CPM) \cite{cpm_traag2011}.
Alternatively, modularity also admits a spectral formulation $\mathcal{Q} = \mathrm{Tr}(\mathbf{C}^\top \mathbf{M} \mathbf{C})/2m$, where $\mathbf{M} = \mathbf{A} - \gamma ({\mathbf{d}_{\mathrm{out}}\mathbf{d}_{\mathrm{in}}^\top})/{2m}$ is the modularity matrix.
In this case, projecting (`embedding') nodes onto the leading eigenvectors of $\mathbf{M}$ yields a low-dimensional representation that approximates the discrete optimization problem, which may be subsequently clustered to recover partitions \cite{spectral_modularity_newman2006} --- for bipartitioning, the sign of the leading eigenvector determines the split, and the absence of positive eigenvalues signals that the current bipartition is optimal under modularity.

The measure has been extended to dynamic graphs through, e.g., the multislice formulation \cite{modularity_mucha2010}, which couples snapshot-level modularity matrices through inter-slice edges encoding temporal dependencies, enabling community detection and tracking across discrete-time aggregations.
Continuous-time extensions, such as the longitudinal formulation \cite{modularity_longitudinal2025}, instead impose smoothness regularization and null model constraints.
Both variants increase the problem size with the number of snapshots or pairwise interactions, compounding the scalability challenges of the static case.
The multislice formulation is compatible with existing optimization methods, such as Leiden, which we efficiently implement on GPUs by building the supra-adjacency matrix encoding intra-slice and inter-slice connections.

Multislice modularity $Q_\mathrm{MS}$ \cite{modularity_mucha2010} is defined analogously to the static case, with additional inter-slice couplings of (optionally time-varying) strength $\omega$ that encourage temporal persistence of community labels across snapshots.
This yields a similar objective function to be maximized, which may be expressed as
\begin{equation}
    \label{eq:multislice-modularity}
    Q_\mathrm{MS} =
    \frac{1}{2\mu}
    \sum_{i,j,s,r}
    \left[
        \left(
            A_{ij}^{(s)} - \gamma^{(s)} \frac{k_i^{(s)} k_j^{(s)}}{2m^{(s)}}
        \right)
        \delta^{(sr)}
        + \delta_{ij}\,\omega^{(sr)}
    \right]
    \delta(c_i^{(s)}, c_j^{(r)}),
\end{equation}
where $A_{ij}^{(s)}$ is the adjacency matrix of snapshot $s$, $k_i^{(s)}$ is the degree of node $i$ in snapshot $s$, $m^{(s)}$ is the number of edges in snapshot $s$, $\gamma^{(s)}$ is the resolution parameter for snapshot $s$, $\omega^{(sr)}$ is the inter-slice node coupling strength between $s$ and $r$, and $\mu$ is the supra-graph size (weighted intra- and inter-slice connections).

Following Equation~\ref{eq:multislice-modularity}, we may express multislice modularity in spectral form by defining a supra-modularity matrix incorporating both intra-slice and inter-slice contributions, and a community assignment matrix $\bm{\mathcal{C}}$ encoding the partition across all snapshots.
This yields a spectral formulation expressed as
$
\mathcal{Q}_{\mathrm{MS}} =
\mathrm{Tr} (\bm{\mathcal{C}}^\top \bm{\mathcal{M}} \bm{\mathcal{C}}) / 2 \mu,
$
where $\bm{\mathcal{M}}$ is the (temporal) modularity matrix with diagonal blocks corresponding to the modularity matrices of each snapshot and off-diagonal blocks encoding the inter-slice couplings, and each block in $\bm{\mathcal{C}} \in \mathbb{R}^{nT \times k}$ corresponds to the community assignments of a snapshot.
If $\bm{\mathcal{C}}$ is binary with one non-zero entry per row, then $Q_{\mathrm{MS}} = \mathcal{Q}_{\mathrm{MS}}$; solving its continuous relaxation approximates the discrete optimization problem, allowing for efficient community detection and tracking across time through eigendecomposition and subsequent clustering of the resulting embeddings (leading eigenvectors) with, e.g., $k$-means.

This spectral relaxation therefore enables the application of sparse eigensolver implementations: unlike the non-backtracking operator, the modularity matrix yields a real, orthogonal eigenvector embedding when symmetric, i.e., for undirected graphs with inter-slice couplings $\omega^{(sr)} = \omega^{(rs)}$.
Meanwhile, the Leiden algorithm and its GPU backends let us either optimize a global null model (Eq.~\ref{eq:modularity}) on the supra-graph as a proxy for $Q_{\mathrm{MS}}$ \cite{cugraph_docs}, or efficiently implement the per-slice multislice null model with specialized data structures and parallelized computations \cite{cupy_learningsys2017}.
Note that the supra-graph optimization strategy is also compatible with alternative quality functions beyond modularity, such as a temporal (multislice) variant of CPM \cite{cpm_traag2011}, provided the null model is appropriately defined.

In sum, both spectral decomposition and greedy optimization allow leveraging GPU acceleration, offering a unified framework for community detection on dynamic graphs.
It is worth noting that the spectral pathway, which directly captures the global structure of the network through its eigenvectors, allows for a more principled approach to detecting communities.
While Leiden optimization relies on heuristics that may not guarantee optimal partitions and can be sensitive to the resolution parameter $\gamma$ and inter-slice coupling strength $\omega$,
it allows for multi-GPU \cite{dask_docs,cugraph_docs} and specialized \cite{cupy_learningsys2017} approaches, especially advantageous for large-scale dynamic graphs, making it a practical choice for diverse applications.

\section{Experimental Results} \label{sec:results}

\begin{table}[t!]
\caption{\textbf{Dataset properties.}
$|\mathcal{V}|$ and $|\mathcal{E}|$ denote unique nodes and edges; $\mathrm{\Sigma}({\mathcal{V}})$ and $\mathrm{\Sigma}({\mathcal{E}})$, the sum of nodes and edges over time, i.e., counting each occurrence across snapshots; $|T|$ and $|C|$ are the number of snapshots and communities; $\dagger$ marks static graphs used as single-snapshot baselines; and $\ddagger$, the largest dataset, where a single CPU run allowed to exceed the time budget completed in $\approx$6\,h, versus $\approx$10\,min on GPU \cite{cugraph_docs}.}
\label{tab:datasets}
\centering
\scriptsize
\setlength{\tabcolsep}{3.8pt}
\renewcommand{\arraystretch}{1.05}
\begin{tabularx}{0.96\linewidth}{@{\hspace{0.6em}}l
    >{\hsize=1.0\hsize\raggedleft\arraybackslash}X
    >{\hsize=1.2\hsize\raggedleft\arraybackslash}X
    >{\hsize=1.3\hsize\raggedleft\arraybackslash}X
    >{\hsize=1.3\hsize\raggedleft\arraybackslash}X
    >{\hsize=0.7\hsize\raggedleft\arraybackslash}X
    >{\hsize=0.5\hsize\raggedleft\arraybackslash}X
@{\hspace{0.6em}}}
\toprule
\textbf{Dataset}
    & $|\mathcal{V}|$
    & $\mathrm{\Sigma}({\mathcal{V}})$
    & $|\mathcal{E}|$
    & $\mathrm{\Sigma}({\mathcal{E}})$
    & $|T|$
    & $|C|$ \\
\midrule
CiteSeer$^{\dagger }$\cite{planetoid_zhiling2016} & 3\,279 & 3\,279   & 4\,552       & 4\,552       & 1      &  6   \\
Cora$^{\dagger }$\cite{planetoid_zhiling2016}     & 2\,708 & 2\,708   & 5\,278       & 5\,278       & 1      &  7   \\
\midrule
Brain$\;$\cite{brain_pretti2017}      & 5\,000      & 60\,000     & 878\,207     & 947\,744     & 12     & 10   \\
Dblp$\;$\cite{dblp_zuo2018}       & 28\,085     & 101\,797    & 153\,822     & 222\,165     & 27     & 10   \\
Patent$\;$\cite{patent_hall2001}     & 12\,214     & 41\,529     & 41\,916      & 41\,916      & 891    &  6   \\
PubMed$\;$\cite{passos2024_gnnet}     & 19\,717     & 37\,003     & 44\,324      & 44\,324      & 42     &  3   \\
School$\;$\cite{school_mastrandrea2015}    & 327         & 309\,006    & 7\,004       & 188\,508     & 7\,375 &  9   \\
\midrule
ArxivAI$\;$\cite{data4tgc2023}    & 69\,854     & 142\,316    & 696\,819     & 696\,819     & 27     &  5   \\
ArxivCS$\;$\cite{data4tgc2023}    & 169\,343    & 374\,433    & 1\,157\,799  & 1\,157\,799  & 29     & 40   \\
ArxivLarge$^\ddagger$\cite{data4tgc2023} & 1\,324\,064 & 4\,998\,391 & 13\,649\,351 & 13\,649\,351 & 40     & 172  \\
ArxivMath$\;$\cite{data4tgc2023}  & 270\,013    & 614\,578    & 783\,165     & 783\,165     & 31     & 31   \\
\midrule
SBM$\;$\cite{tadcsbm2025} & 50\,000  & 250\,000 & 1\,250\,539 & 1\,251\,057 & 5  & 10 \\
SBM-N$\;$\cite{tadcsbm2025} & 100\,000 & 500\,000 & 2\,500\,515 & 2\,501\,000 & 5  & 10 \\
SBM-E$\;$\cite{tadcsbm2025} & 50\,000  & 250\,000 & 2\,498\,067 & 2\,500\,045 & 5  & 10 \\
SBM-T$\;$\cite{tadcsbm2025} & 50\,000  & 500\,000 & 2\,495\,376 & 2\,497\,636 & 10 & 10 \\
\bottomrule
\end{tabularx}
\vspace{-1.0em}
\end{table}

\begin{figure*}[t!]
\centering
\includegraphics[width=\textwidth]{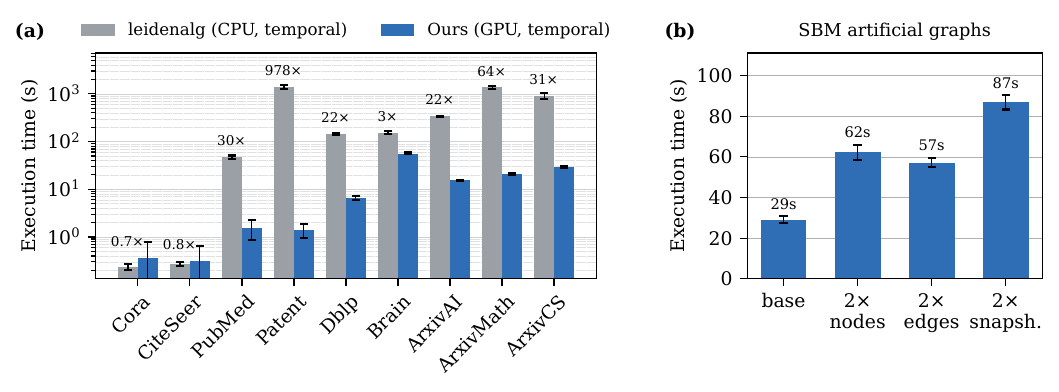}\\[-1em]
\caption{\textbf{Runtime evaluation.}
(a) Execution time of the GPU-based Leiden backend \cite{cugraph_docs} against its CPU-bound baseline across real-world datasets (log scale).
(b) Effect on runtime of doubling each generation parameter (nodes, edges, snapshots) over synthetic SBM instances.
Algorithm runs were limited to two iterations only (\texttt{leidenalg} \cite{leidenalg_docs} default) to ensure fair comparisons.
Results reported over five runs, including host-device transfer and kernel launch overheads; error bars show standard deviation.
Experiments conducted on an Intel Xeon Gold 6330 (CPU) and an NVIDIA A100 80GB (GPU).}
\label{fig:results}
\end{figure*}

We present experimental results of our code implementation on several real-world and synthetic dynamic graph datasets, summarized in Table~\ref{tab:datasets} and Figure~\ref{fig:results}.

Our central comparison is temporal-versus-temporal: the GPU Leiden backend \cite{cugraph_docs}, optimizing a global null model on the supra-graph, against its multislice CPU reference \cite{leidenalg_docs}.
For fairness across implementations with differing stopping criteria, all methods are capped at two optimization passes; we therefore report runtime under a fixed, equal-work budget rather than time-to-convergence.
Observed speedups (Fig.~\ref{fig:results}a) are consistent and substantial, ranging from $22\times$ on \textit{ArxivAI} (336.3\,s\,$\rightarrow$\,15.4\,s) and \textit{Dblp} (145.5\,s\,$\rightarrow$\,6.6\,s) to $31\times$ on \textit{ArxivCS} (916.3\,s\,$\rightarrow$\,29.2\,s) and $64\times$ on \textit{ArxivMath} (1357.7\,s\,$\rightarrow$\,21.2\,s); dense but low-snapshot \textit{Brain} yields a more modest $2.7\times$ (154.7\,s $\rightarrow$ 56.3\,s).
The most striking case is \textit{Patent}, collapsing by $978\times$ (1397.0\,s $\rightarrow$ 1.4\,s).
This is precisely the scaling challenge our work targets, and constitutes our main empirical result, as it demonstrates the practical benefits of GPU acceleration for fine-grained community detection on datasets with a large number of snapshots, where CPU-bound methods struggle to meet demand.

For the largest temporal instances, the effect is less a speedup than a change in tractability (Tab.~\ref{tab:datasets}, $\ddagger$).
We further report on two interesting boundary cases.
First, \textit{School} (7375 snapshots) exposes a scaling limit of the supra-graph construction: the inter-slice coupling grows with the snapshot count, so building and solving the supra-adjacency dominates (211\,s for our method), with the CPU baseline failing to complete within the time budget (OOT);
this limit motivates approximate strategies such as windowed optimization or distributed sparse construction.
Second, the GPU advantage emerges only at scale: for small static graphs (\textit{CiteSeer}, \textit{Cora}), the CPU outperforms the GPU backend, as kernel launch and transfer overheads dominate;
these could potentially be mitigated by assembling the supra-graph directly on device.
Moreover, the synthetic SBM graphs (Fig.~\ref{fig:results}b) characterize how the GPU implementation scales along distinct axes (nodes, edges, snapshots) rather than draw head-to-head comparisons against the CPU.

\section{Final Remarks}

In this work, we presented GPU-accelerated implementations of spectral clustering and modularity optimization for dynamic graphs, to our knowledge the first in literature to leverage the RAPIDS ecosystem for efficient matrix operations and eigensolvers on GPU hardware; and a unified integration of both as a zero-code-change backend for NetworkX-Temporal, yielding concrete usability benefits.
Our implementations are designed to be compatible with existing graph analytics pipelines, allowing for seamless integration with data processing workflows without extensive codebase modifications.
Limitations include the focus on non-attributed graphs, which may not capture the full richness of real-world networks, and the reliance on assortative community structures, which is not suitable for all contexts.
Lastly, we leave evaluating the empirical performance of our spectral and parallelized approaches under different detectability regimes to future work.


\begingroup
\let\small\footnotesize
\renewcommand{\doi}[1]{}
\renewcommand{\url}[1]{}
\bibliographystyle{splncs04}
\bibliography{references}
\endgroup

\end{document}